\makeatletter
\declare@file@substitution{revtex4-1.cls}{revtex4-2.cls}
\makeatother

\documentclass[twocolumn,tighten]{aastex631}
\hypersetup{linkcolor=red,citecolor=blue,filecolor=cyan,urlcolor=magenta}

\usepackage[fleqn]{amsmath}
\usepackage{natbib}
\usepackage{amssymb}
\usepackage{txfonts}
\usepackage{graphicx}
\usepackage{multirow}
\usepackage{dcolumn}
\usepackage{array}
\usepackage{xcolor}
\usepackage{float}
\usepackage{bm}

\received{MM DD, YY}
\revised{MM DD, YY}
\accepted{MM DD, YY}
\shorttitle{Probe the regolith characteristics of asteroids}
\shortauthors{Yu L.L.}

\begin{document}
\title{Probe the regolith characteristics of asteroids from 9-years infrared observations of WISE/NEOWISE: A case study of the
Main-Belt Object (656) Beagle}

\correspondingauthor{Liang-Liang Yu}
\email{yullmoon@nju.edu.cn}

\author{Liang-Liang Yu}
\affiliation{Institute of Science and Technology for Deep Space Exploration, Nanjing University-Suzhou Campus, Suzhou, 215163, China}

\begin{abstract}
This work presents data processing, fitting procedure, modelling and analyzing of 9-years infrared light curves provided by the WISE/NEOWISE telescope, by which the regolith characteristics of Main-Belt Object (656) Beagle is studied. 
We determine Beagle's effective diameter $D_{\rm eff}=57.3^{+4.5}_{-2.2}$ km, geometric albedo $p_{\rm v}=0.05^{+0.004}_{-0.007}$, mean roughness $\theta_{\rm RMS}=44\pm4^\circ$, mean grain size $b=100^{+350}_{-90}~\mu$m, mean specific heat capacity $c_{\rm p}=173\sim516\rm~JKg^{-1}K^{-1}$, mean thermal conductivity $\kappa=0.7\sim1.3\times10^{-3}\rm~Wm^{-1}K^{-1}$ and mean thermal inertia $\Gamma=14\sim32\rm~Jm^{-2}s^{-0.5}K^{-1}$. 
The albedo of Beagle is a little anomalous that the albedos of Beagle's neighbouring asteroids are more close to Themis, rather than Beagle itself. The W1-band near-infrared light curves don't reveal significant heterogeneous NIR features on the surface of Beagle, being inconsistent with the expectation of a family parent that has members with diverse NIR spectral types. These results add new clues of Beagle probably being an interloper or a sister, rather than the parent of its neighbouring asteroids including the first main-belt comet (MBC) 133P, hence may lead to new scenarios about the origin of famous MBC 133P. Besides, we found that asteroidal shape models from inversion of optical light curves are imperfect for modeling infrared lightcurves, thus could mislead evaluations of both the heterogeneity of regolith reflectivity at near infrared and thermophysical characteristics at thermal infrared.
\end{abstract}


\section{Introduction}
The main-belt asteroid (656) Beagle (hereafter Beagle for short) came into the sight
of astronomers since it was proposed to be the parent of the famous main-belt comet (MBC)
133P/Elst-Pizarro \citep{Nesvorny2008}. And the Beagle family is proposed to be a young
family with an age of $<10$ Myr \citep{Nesvorny2008} or $\sim14$ Myr \citep{Carruba2019}.
Then the story about the origin of 133P is widely accepted as the following scenario:
133P is a young icy fragment of Beagle, which is a daughter of the icy main-belt object 
(24) Themis, on which absorption feature of water ice was observed by the IRTF spectra 
\citep{Campins2010,Rivkin2010}. The formation scenario was somewhat supported by the work of 
\citep{Fornasier2016}, which reported Beagle to have a geometric albedo of $p_{\rm v}=0.0782\pm0.0222$, 
being similar to the geometric albedos of its neighbouring asteroids (mean $p_{\rm v}=0.0941\pm0.0055$) 
and potential parent (24) Themis ($p_{\rm v}=0.0641\pm0.0157$). 

Besides, \citet{Fornasier2016} also found that the neighbouring asteroids of Beagle exhibit 
diverse spectral types based on the reflectance spectrum at near infrared (NIR), implying that 
the composition of these asteroids differs from each other. If these neighbouring asteroids 
of Beagle were formed from the same impact event, then their diverse compositions may originate 
from the parent, or from the impactor, or produced by the high temperature and pressure during 
impact process. According to \citet{Michel2015}, if a parent body is heterogeneous in composition, 
then the resulting family is expected to show a variety of spectral properties within its members. 
On the other hand, although the original parent may not have spectral variability on its 
surface, the impact that produced heterogenous members will make the survived parent to show 
surface heterogeneity. For example, Vesta is observed to show heterogeneous surface features 
due to impact exposing heterogeneous subsurface and upper crust materials \citep{Rousseau2021}. 
Therefore, if Beagle is the survived parent of these neighbouring asteroids with diverse compositions, 
then Beagle is expected to contain diverse compositions on its surface and show heterogeneous NIR 
features across its surface.

However, \citet{Mainzer2016} reported Beagle to have a very low geometric albedo 
$p_{\rm v}=0.045\pm0.005$, being even lower than that of (24) Themis. On the other hand, 
the recent work of \citet{Yu2020} found that 133P is more likely to have an old age $>100$ Myr, 
being inconsistent with the young age ~$<10$ Myr of the Beagle family. These new observations 
and theoretical predictions thus raise the question of whether Beagle having genetic connections 
with 133P and (24) Themis. 

So in this work, we attempt to use more data and new method to investigate whether asteroid Beagle 
has a geometric albedo as low as $\sim0.05$ and significant heterogeneous reflectivity at near 
infrared. This objective can be achieved by analyzing the multi-year infrared light curves 
of Beagle from WISE/NEOWISE with the well-tested thermophysical model --- RSTPM \citep{Yu2021}.  
As the WISE/NEOWISE observations of main-belt objects at band-W1 are dominated by sunlight reflection 
at near infrared. In addition, heterogeneity of regolith thermophysical characteristics can also be 
evaluated from the other three bands of W2, W3, and W4. 

\section{The Radiometric model}
\subsection{Observations and data processing}
The {\it Wide-field Infrared Survey Explorer} (WISE) mission has mapped entire sky in
four bands around 3.4 (W1), 4.6 (W2), 12 (W3), and 22 (W4) $\mu$m with resolutions from
$6.1^{\prime\prime}$ to $12^{\prime\prime}$ \citep{Wright2010}. All four bands were imaged 
simultaneously, and the exposure times were 7.7 s in W1 and W2, and 8.8 s in W3 and W4 
\citep{Wright2010}. The four-bands survey started from 2010 January 7, and ended on 2010 
August 6 after the outer cryogen tank was exhausted, making the W4 channel be no longer 
able to be used to obtain survey data. The W3 channel continued operation until 2010 
September 29 when the inner cryogen reserve was exhausted, while the W1 and W2 channel 
kept working until the telescope was set into hibernation on 2011 February 1 
\citep{Mainzer2011,Cutri2012}. The two-band survey was then resumed on 2013 December 
13 (known as NEOWISE) \citep{Mainzer2014,Cutri2015};, and is still in service, which
has obtained nearly 10-year observations.

We found 9-years observations of Beagle from the WISE archive (see the website of the
NASA/IPAC Infrared Science Archive http://irsa.ipac.caltech.edu/). The datasets are 
summerized in Table \ref{obs0}. 
\begin{table}[htbp]
\centering
\renewcommand\arraystretch{1.1}
\caption{The collected WISE/NEOWISE observations of (656) Beagle from 2010 to 2019.}
\label{obs0}
\begin{tabular}{@{}cccccc@{}}
\hline
 UTC & Bands & Obs & $r_{\rm helio}$ & $\Delta_{\rm obs}$ & $\alpha$  \\
     &       & Num & (AU) & (AU) & ($^{\circ}$)  \\
\hline
Jan. 31-01, 2010 & W1-4 & 14 & 2.82 & 2.66 & 20.42 \\
Jul. 26-28, 2010 & W1-4 & 16 & 3.03 & 2.78 & -19.51 \\
Feb. 15-16, 2014 & W1-2 & 11 & 3.01 & 2.81 & -19.17 \\
Dem. 05-06, 2014 & W1-2 & 14 & 2.73 & 2.52 & 21.11 \\
May. 19-20, 2015 & W1-2 & 16 & 2.75 & 2.40 & -21.28 \\
Mar. 09-11, 2016 & W1-2 & 11 & 3.06 & 2.88 & 18.95 \\
Aug. 20-21, 2016 & W1-2 & 11 & 3.27 & 2.88 & -17.52 \\
May. 14-15, 2017 & W1-2 & 14 & 3.52 & 3.39 & 16.70 \\
Oct. 19-20, 2017 & W1-2 & 12 & 3.57 & 3.12 & -15.32 \\
Jul. 17-18, 2018 & W1-2 & 11 & 3.49 & 3.33 & 16.95 \\
Dem. 12-13, 2018 & W1-2 & 13 & 3.35 & 2.84 & -15.66 \\
Oct. 03-04, 2019 & W1-2 & 12 & 2.99 & 2.81 & 19.58 \\
\hline
\multicolumn{6}{l}{$r_{\rm helio}$: heliocentric distance;} \\
\multicolumn{6}{l}{$\Delta_{\rm obs}$: observation distance;} \\
\multicolumn{6}{l}{$\alpha$: solar phase angle.} \\
\end{tabular}
\end{table}

\subsubsection{Flux colour corrections}
All the four-band data need colour corrections, especially for the W1 and W2-band data, 
the infrared fluxes of which contain not only self-thermal emission but also sunlight 
diffused from the object’s surface, thus the color corrections should be done separately 
for the thermal component and the reflection component, as the color correction factors 
are different. Besides, the color correction factor for the thermal component is temperature 
dependent, thus will have different value if the helio-centric distance was different 
for each observation epoch. So we implement the color-correction procedure as follows: 

First, convert the data-base magnitude to flux without any color correction, giving the total 
integrated flux $F(\lambda\pm\Delta\lambda)_{\rm tot,obs}$ for each band. The derived band-integrated 
fluxes together with the observation geometry are listed in Table \ref{obs1} and \ref{obs2}, and 
each flux should at least have an associated uncertainty of $\pm$10 percent according to \citet{Wright2010}; 

\begin{table}[htbp]\footnotesize
\centering
\renewcommand\arraystretch{0.9}
\caption{WISE observations of (656) Beagle. The derived fluxes are band-integrated fluxes of each band without color correction. Each flux should have an associated uncertainty of ±10 percent according to \citet{Wright2010}.}
\label{obs1}
\begin{tabular}{@{}cccccccc@{}}
\hline
 MJD & W1 & W2 & W3 & W4 & $r_{\rm helio}$ & $\Delta_{\rm obs}$ & $\alpha$  \\
     & (mJy) & (mJy) & (Jy) & (Jy) & (AU) & (AU) & ($^{\circ}$)  \\
\hline
55227.504 & 1.49 & 6.64 & 1.26 & 3.18 & 2.82 & 2.67 & 20.42 \\
55227.637 & 1.44 & 6.71 & 1.16 & 2.99 & 2.82 & 2.67 & 20.42 \\
55227.637 & 1.51 & 6.57 & 1.22 & 3.22 & 2.82 & 2.67 & 20.42 \\
55227.769 & 1.42 & 5.44 & 0.94 & 2.58 & 2.82 & 2.67 & 20.42 \\
55227.901 & 0.96 & - & 0.74 & 2.16 & 2.82 & 2.66 & 20.42 \\
55228.034 & 0.77 & 2.95 & 0.54 & 1.64 & 2.82 & 2.66 & 20.42 \\
55228.100 & 1.53 & 6.93 & 1.16 & 3.37 & 2.82 & 2.66 & 20.42 \\
55228.166 & 0.73 & 2.84 & 0.57 & 1.69 & 2.82 & 2.66 & 20.42 \\
55228.232 & 1.58 & 7.36 & 1.23 & 3.34 & 2.82 & 2.66 & 20.42 \\
55228.298 & 0.87 & 4.14 & 0.75 & 2.25 & 2.82 & 2.66 & 20.42 \\
55228.364 & 1.29 & 5.95 & 1.06 & 2.93 & 2.82 & 2.66 & 20.42 \\
55228.497 & 1.17 & 4.68 & 0.87 & 2.44 & 2.82 & 2.66 & 20.42 \\
55228.629 & 0.95 & 3.90 & 0.73 & 1.95 & 2.82 & 2.65 & 20.42 \\
55228.762 & - & 3.23 & 0.55 & 1.59 & 2.82 & 2.65 & 20.42 \\
55403.908 & 1.17 & 3.00 & 0.68 & 2.18 & 3.03 & 2.77 & -19.50 \\
55404.040 & 0.80 & - & 0.50 & 1.70 & 3.03 & 2.77 & -19.50 \\
55404.172 & 0.58 & 1.73 & 0.40 & 1.34 & 3.03 & 2.77 & -19.51 \\
55404.173 & 0.58 & 1.68 & 0.41 & 1.34 & 3.03 & 2.77 & -19.51 \\
55404.305 & 0.62 & 1.59 & 0.36 & 1.19 & 3.03 & 2.77 & -19.51 \\
55404.371 & 1.21 & 3.68 & 0.88 & 2.54 & 3.03 & 2.78 & -19.51 \\
55404.437 & 0.76 & 2.05 & 0.41 & 1.30 & 3.03 & 2.78 & -19.51 \\
55404.503 & 1.01 & 2.70 & 0.70 & 2.32 & 3.03 & 2.78 & -19.51 \\
55404.569 & 0.90 & 2.67 & 0.58 & 1.76 & 3.03 & 2.78 & -19.51 \\
55404.636 & 0.98 & 2.33 & 0.63 & 2.01 & 3.03 & 2.78 & -19.51 \\
55404.768 & 0.81 & 2.02 & 0.48 & 1.63 & 3.03 & 2.78 & -19.51 \\
55404.900 & 0.38 & - & 0.35 & 1.20 & 3.03 & 2.78 & -19.51 \\
55405.032 & 0.50 & 1.43 & 0.33 & 1.02 & 3.03 & 2.79 & -19.52 \\
55405.032 & 0.53 & 1.26 & 0.33 & 1.04 & 3.03 & 2.79 & -19.52 \\
55405.165 & 0.68 & 2.09 & 0.45 & 1.41 & 3.03 & 2.79 & -19.52 \\
55405.165 & 0.65 & 2.31 & 0.50 & 1.53 & 3.03 & 2.79 & -19.52 \\
\hline
\end{tabular}
\end{table}

\begin{table*}[htbp]\small
\centering
\renewcommand\arraystretch{0.7}
\caption{NEOWISE observations of (656) Beagle. The derived fluxes are band-integrated fluxes of each band without color correction. Each flux should have an associated uncertainty of ±10 percent according to \citet{Wright2010}.} 
\label{obs2}
\begin{tabular}{@{}ccccccccccccc@{}}
\hline
 MJD & W1 & W2  & $r_{\rm helio}$ & $\Delta_{\rm obs}$ & $\alpha$ && MJD & W1 & W2  & $r_{\rm helio}$ & $\Delta_{\rm obs}$ & $\alpha$ \\
     & (mJy) & (mJy) & (AU) & (AU) & ($^{\circ}$)  &&     & (mJy) & (mJy) & (AU) & (AU) & ($^{\circ}$)  \\
\hline
56703.138 & 0.99 & 2.23 & 3.01 & 2.80 & -19.16  & &     57887.678 & 0.32 & 0.61 & 3.52 & 3.40 & 16.70 \\
56703.270 & 0.75 & 1.69 & 3.01 & 2.80 & -19.16  & &     57887.809 & 0.25 & - & 3.52 & 3.40 & 16.70 \\
56703.401 & 0.54 & 1.21 & 3.01 & 2.80 & -19.16  & &     57887.940 & 0.36 & 0.65 & 3.52 & 3.39 & 16.70 \\
56703.533 & 0.66 & 1.68 & 3.01 & 2.81 & -19.17  & &     57887.940 & 0.41 & 0.52 & 3.52 & 3.39 & 16.70 \\
56703.599 & 1.18 & 3.22 & 3.01 & 2.81 & -19.17  & &     57888.071 & 0.54 & 0.80 & 3.52 & 3.39 & 16.70 \\
56703.665 & 1.01 & 2.80 & 3.01 & 2.81 & -19.17  & &     57888.136 & 0.47 & 0.67 & 3.52 & 3.39 & 16.70 \\
56703.731 & 1.14 & 2.62 & 3.01 & 2.81 & -19.17  & &     57888.202 & 0.64 & 0.86 & 3.52 & 3.39 & 16.70 \\
56703.797 & 1.05 & 3.13 & 3.01 & 2.81 & -19.17  & &     57888.267 & 0.34 & 0.47 & 3.52 & 3.39 & 16.70 \\
56703.994 & 0.60 & 1.36 & 3.01 & 2.81 & -19.18  & &     57888.333 & 0.68 & 1.23 & 3.52 & 3.39 & 16.71 \\
56704.126 & 0.70 & 1.54 & 3.01 & 2.81 & -19.18  & &     57888.333 & 0.71 & 1.38 & 3.52 & 3.39 & 16.71 \\
56704.258 & 0.72 & 2.64 & 3.01 & 2.82 & -19.18  & &     57888.398 & 0.28 & 0.51 & 3.52 & 3.39 & 16.71 \\
56996.277 & 1.71 & 8.14 & 2.73 & 2.53 & 21.12   & &     57888.529 & 0.36 & 0.45 & 3.52 & 3.39 & 16.71 \\
56996.409 & 1.51 & 7.25 & 2.73 & 2.53 & 21.12   & &     57888.660 & 0.56 & 0.82 & 3.52 & 3.38 & 16.71 \\
56996.540 & 1.17 & 5.58 & 2.73 & 2.53 & 21.11   & &     57888.791 & 0.60 & 0.98 & 3.52 & 3.38 & 16.71 \\
56996.672 & 0.84 & 3.77 & 2.73 & 2.52 & 21.11   & &     58045.491 & 0.40 & 0.53 & 3.57 & 3.12 & -15.28 \\
56996.737 & 1.91 & 9.87 & 2.73 & 2.52 & 21.11   & &     58045.622 & 0.38 & 0.46 & 3.57 & 3.12 & -15.29 \\
56996.803 & 0.66 & 3.16 & 2.73 & 2.52 & 21.11   & &     58045.622 & 0.31 & 0.53 & 3.57 & 3.12 & -15.29 \\
56996.869 & 1.84 & 9.27 & 2.73 & 2.52 & 21.11   & &     58045.753 & 0.44 & 0.59 & 3.57 & 3.12 & -15.30 \\
56996.934 & 0.94 & 4.62 & 2.73 & 2.52 & 21.11   & &     58045.884 & 0.56 & 0.97 & 3.57 & 3.12 & -15.31 \\
56997.000 & 1.64 & 8.15 & 2.73 & 2.52 & 21.11   & &     58046.015 & 0.57 & 1.06 & 3.57 & 3.12 & -15.32 \\
56997.066 & 1.53 & 7.41 & 2.73 & 2.52 & 21.11   & &     58046.080 & 0.50 & 0.69 & 3.57 & 3.13 & -15.32 \\
56997.132 & 1.37 & 6.12 & 2.73 & 2.52 & 21.11   & &     58046.146 & 0.93 & 1.08 & 3.57 & 3.13 & -15.33 \\
56997.263 & 0.93 & 4.10 & 2.73 & 2.52 & 21.11   & &     58046.211 & 0.39 & 0.42 & 3.57 & 3.13 & -15.34 \\
56997.394 & 0.76 & 3.28 & 2.73 & 2.51 & 21.11   & &     58046.342 & 0.39 & 0.54 & 3.57 & 3.13 & -15.35 \\
56997.526 & 0.96 & 4.15 & 2.73 & 2.51 & 21.10   & &     58046.473 & 0.52 & 0.57 & 3.57 & 3.13 & -15.36 \\
57161.421 & 1.82 & 7.32 & 2.75 & 2.39 & -21.25  & &     58046.604 & 0.67 & 0.94 & 3.57 & 3.13 & -15.37 \\
57161.552 & 1.41 & 5.31 & 2.75 & 2.40 & -21.26  & &     58316.418 & 0.33 & 0.60 & 3.49 & 3.34 & 16.95 \\
57161.552 & 1.65 & 5.52 & 2.75 & 2.40 & -21.26  & &     58316.549 & 0.44 & 0.84 & 3.49 & 3.34 & 16.95 \\
57161.683 & 1.20 & 3.57 & 2.75 & 2.40 & -21.26  & &     58316.680 & 0.57 & 0.95 & 3.49 & 3.34 & 16.95 \\
57161.814 & 0.87 & 2.76 & 2.75 & 2.40 & -21.27  & &     58316.811 & 0.67 & 1.26 & 3.49 & 3.33 & 16.95 \\
57161.880 & 1.87 & 7.80 & 2.75 & 2.40 & -21.27  & &     58316.942 & 0.62 & 1.20 & 3.49 & 3.33 & 16.95 \\
57161.946 & 0.89 & 3.42 & 2.75 & 2.40 & -21.28  & &     58317.007 & 0.36 & 0.53 & 3.49 & 3.33 & 16.95 \\
57161.946 & 0.87 & 3.51 & 2.75 & 2.40 & -21.28  & &     58317.073 & 0.63 & 1.03 & 3.49 & 3.33 & 16.95 \\
57162.011 & 1.98 & 8.11 & 2.75 & 2.40 & -21.28  & &     58317.138 & 0.37 & 0.71 & 3.49 & 3.33 & 16.95 \\
57162.077 & 1.22 & 5.61 & 2.75 & 2.40 & -21.28  & &     58317.269 & 0.59 & 0.85 & 3.49 & 3.33 & 16.95 \\
57162.143 & 1.63 & 5.65 & 2.75 & 2.40 & -21.29  & &     58317.400 & 0.72 & 1.18 & 3.49 & 3.32 & 16.95 \\
57162.274 & 1.19 & 4.16 & 2.75 & 2.41 & -21.29  & &     58317.531 & 0.57 & 0.97 & 3.49 & 3.32 & 16.95 \\
57162.405 & 0.95 & 3.04 & 2.75 & 2.41 & -21.30  & &     58464.862 & 0.47 & 0.90 & 3.35 & 2.83 & -15.60 \\
57162.536 & 1.02 & 3.38 & 2.75 & 2.41 & -21.31  & &     58464.992 & 0.62 & 1.29 & 3.35 & 2.84 & -15.62 \\
57162.536 & 0.89 & 3.22 & 2.75 & 2.41 & -21.31  & &     58464.992 & 0.71 & 1.27 & 3.35 & 2.84 & -15.62 \\
57162.667 & 1.04 & 5.02 & 2.75 & 2.41 & -21.31  & &     58465.123 & 0.86 & 1.63 & 3.35 & 2.84 & -15.63 \\
57456.914 & 1.39 & 3.76 & 3.06 & 2.89 & 18.95   & &     58465.254 & 1.03 & 1.87 & 3.35 & 2.84 & -15.65 \\
57457.045 & 1.02 & 3.40 & 3.06 & 2.89 & 18.95   & &     58465.320 & 0.55 & 0.77 & 3.35 & 2.84 & -15.66 \\
57457.176 & 1.19 & 3.26 & 3.06 & 2.89 & 18.95   & &     58465.385 & 1.13 & 2.01 & 3.35 & 2.84 & -15.66 \\
57457.307 & 0.98 & 2.63 & 3.06 & 2.89 & 18.95   & &     58465.450 & 0.40 & - & 3.35 & 2.84 & -15.67 \\
57457.373 & 1.24 & 2.93 & 3.06 & 2.89 & 18.95   & &     58465.451 & 0.42 & 0.78 & 3.35 & 2.84 & -15.67 \\
57457.438 & 0.93 & 2.10 & 3.06 & 2.88 & 18.95   & &     58465.516 & 0.97 & 1.61 & 3.35 & 2.84 & -15.68 \\
57457.635 & - & 3.98 & 3.06 & 2.88 & 18.94      & &     58465.581 & 0.67 & 1.19 & 3.35 & 2.84 & -15.69 \\
57457.766 & 1.18 & 3.48 & 3.06 & 2.88 & 18.94   & &     58465.712 & 0.84 & 1.45 & 3.35 & 2.84 & -15.70 \\
57457.897 & 1.08 & 2.94 & 3.06 & 2.88 & 18.94   & &     58465.843 & 0.98 & 1.83 & 3.35 & 2.85 & -15.72 \\
57458.028 & 0.68 & 2.09 & 3.06 & 2.88 & 18.94   & &     58759.742 & 0.50 & 1.67 & 2.99 & 2.82 & 19.58 \\
57458.159 & 0.68 & 1.62 & 3.06 & 2.88 & 18.94   & &     58759.742 & 0.55 & 1.67 & 2.99 & 2.82 & 19.58 \\
57620.359 & 0.44 & - & 3.27 & 2.88 & -17.49     & &     58759.872 & 0.59 & 2.11 & 2.99 & 2.82 & 19.58 \\
57620.490 & 0.48 & 0.99 & 3.27 & 2.88 & -17.50  & &     58760.134 & 0.92 & 3.20 & 2.99 & 2.81 & 19.58 \\
57620.621 & 0.81 & 1.47 & 3.27 & 2.88 & -17.51  & &     58760.200 & 0.70 & 2.54 & 2.99 & 2.81 & 19.58 \\
57620.752 & 0.92 & 2.06 & 3.27 & 2.88 & -17.52  & &     58760.265 & 1.06 & 4.34 & 2.99 & 2.81 & 19.58 \\
57620.817 & 0.56 & 1.06 & 3.27 & 2.88 & -17.52  & &     58760.330 & 0.76 & 2.04 & 2.99 & 2.81 & 19.58 \\
57620.883 & - & 2.43 & 3.27 & 2.88 & -17.52     & &     58760.396 & 1.21 & 4.10 & 2.99 & 2.81 & 19.58 \\
57620.948 & 0.58 & 0.92 & 3.27 & 2.88 & -17.53  & &     58760.461 & 0.66 & 1.62 & 2.99 & 2.81 & 19.58 \\
57621.014 & 0.84 & 2.00 & 3.27 & 2.89 & -17.53  & &     58760.592 & 0.68 & 2.35 & 2.99 & 2.81 & 19.58 \\
57621.079 & 0.49 & 1.03 & 3.27 & 2.89 & -17.53  & &     58760.723 & 0.90 & 3.02 & 2.99 & 2.80 & 19.58 \\
57621.210 & 0.71 & 1.56 & 3.27 & 2.89 & -17.54  & &     58760.854 & 1.23 & 4.36 & 2.98 & 2.80 & 19.58 \\
57621.341 & 0.76 & 1.93 & 3.27 & 2.89 & -17.55  & &     &&&&& \\ 
\hline
\end{tabular}
\end{table*}

Second, $F(\lambda\pm\Delta\lambda)_{\rm tot,obs}$ contains both integrated thermal emission 
and sunlight reflection, and can be modelled as
\begin{equation}
F(\lambda\pm\Delta\lambda)_{\rm tot,model}= F(\lambda)_{\rm rl}*f_{\rm c,rl}+ F(\lambda)_{\rm th}*f_{\rm c,th}, 
\label{Ftot}
\end{equation}
where $F(\lambda)_{\rm rl}$ and $F(\lambda)_{\rm th}$ represent the monochromatic 
sunlight reflection and thermal emission that can be calculated from reflection 
and thermal models; $f_{\rm c,rl}$ and $f_{\rm c,th}$ stand for the color correction 
factor of the reflection component and the thermal component respectively; here  
$f_{\rm c,rl}$ can be chosen to be the color correction factor of G2V star, 
thus fixing 
\[f_{\rm c,rl}=1.0049,1.0193,1.0024,1.0012\] 
for the four band --- W1, W2, W3 and W4 respectively \citep{Wright2010}; 

Third, since $f_{\rm c,th}$ is temperature dependent, so its value for each band 
will be obtained from an interpolation procedure according to the effective 
temperature 
\begin{equation}
T_{\rm eff}=\left[\frac{(1-A_{\rm B})L_\odot}{\varepsilon\sigma \pi d_\odot^2}\right]^{1/4}
\approx \frac{300\rm~K}{\sqrt{d_\odot}},
\label{Teff}
\end{equation}
of the asteroid at the time of observation, on the basis of the Table 1 of \citet{Wright2010}.
In Equation (\ref{Teff}), $d_\odot$ means the heliocentric distance in AU; $L_\odot$ is the 
Solar constant, about $1361.5\rm~Wm^{-2}$, $A_{\rm B}$ is the surface bond albedo; $\varepsilon$ 
is the averaged thermal emissivity over the entire emission spectrum of the surface and 
$\sigma$ is the Stefan Boltzmann constant; 

Finally, calculate the theoretical monochromatic sunlight reflection $F(\lambda)_{\rm rl}$  
and thermal emission $F(\lambda)_{\rm th}$ via our RSTPM code, and then obtain the theoretical 
total flux $F(\lambda\pm\Delta\lambda)_{\rm tot,model}$ via Equation (\ref{Ftot}), so as to do 
comparison with the observed total flux $F(\lambda\pm\Delta\lambda)_{\rm tot,obs}$ at the first step.

\subsubsection{Anomalous data removal}
Due to some noise pollution, e.g. cosmic rays and zodiacal light, some of the observed fluxes 
may deviate too much from the actual emission of the target. Such anomalous data need to be 
removed from the parameter inversion procedure, so as to make the results about the physical 
characteristics of the target get to be more close to reality.

The basic way to remove anomalous data is on the basis of 'signal to noise ratio ($snr$)'. 
For the WISE/NEOWISE observations of asteroids, generally we remove the data with $snr<3$.

For asteroids with shape model, there is also an additional way to remove anomalous data. 
As the multi-year data of WISE and NEOWISE make it possible to generate infrared light curves, 
so by doing comparisons between the observational and theoretical infrared light curves, 
the data sets deviate too much from the theoretical curves can be treated as anomalous data 
and remove them can make the reversion procedure get better results.

\subsection{Themophysical model}
As has been mentioned above, this work uses the thermophysical model for realistic surface 
layers on airless small bodies --- RSTPM \citep{Yu2021}. The model considers not only real 
shape and rough surface, but also real orbital cycle, rotational cycle, and even temperature 
dependent thermal parameters in the simulation process, as well as contribution of sunlight 
reflection in the infrared radiometric procedure.

In comparison to previous models, the advantages of the RSTPM are: (1) A different mathematical 
technique is used to solve the influence of surface roughness on the energy balance equation of 
surface boundary; (2) For the aim to remove the degeneracy of thermal inertia and roughness by 
interpreting multi-epoch thermal light-curves, variation of thermal parameters due to temperature
variation caused by orbital cycle and rotation cycle is taken into consideration; (3) A combination
model of simultaneously computing thermal emission and sunlight-reflection under the same surface
topography is proposed to fit infrared data in case of the data containing significant sunlight reflection.

So RSTPM has advantages for the small bodies which has regolith on the surface and has large orbital 
eccentricity or an obliquity close to 90 degrees. And for the target asteroid Beagle, we assume it 
have fine regolith on its surface, as it has a much larger size than Eros, Ryugu and Bennu.

\section{Fitting procedure and Results}
\subsection{The Input and Free Parameters}
In order to interpret the multi-year observations, RSTPM needs several input
parameters, including observation geometry, shape model, spin orientation, rotation
phase $ph$, scattering weight-factor $w_{\rm f}$, geometric albedo $p_{\rm v}$, 
effective diameter $D_{\rm eff}$, mean grain radius $b$, and roughness 
fraction $f_{\rm r}$, which can be related to the root-mean-square slope as 
$\theta_{\rm RMS}=\sqrt{f_{\rm r}}\times50^\circ$ \citep{Yu2021}.

\begin{figure}[htbp]
\includegraphics[scale=0.58]{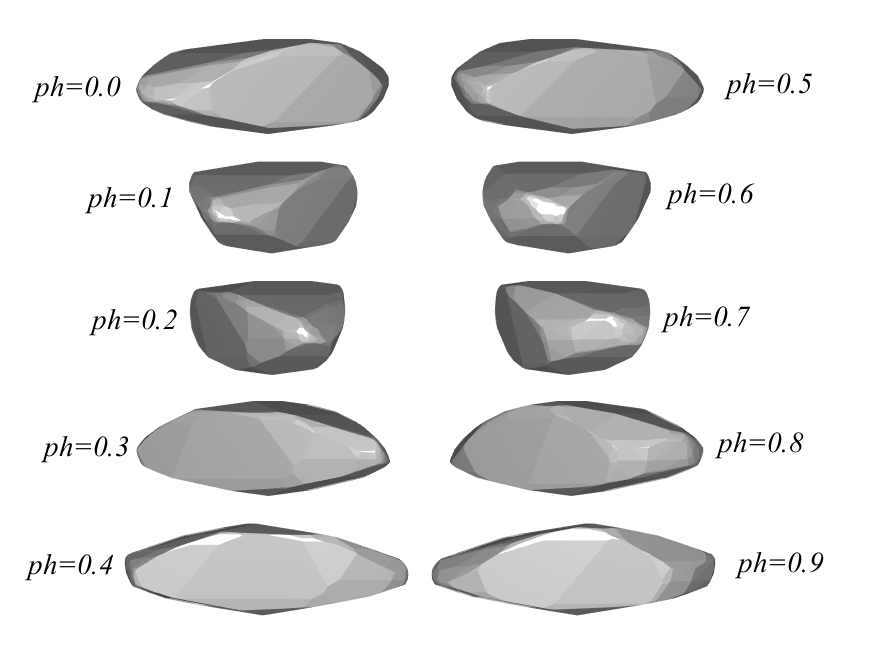}
  \centering
  \caption{The light-curve inversion shape model of (656) Beagle from the DAMIT database. $ph$ represents rotational phase defined as $ph=1-\varphi/(2\pi)$, in which $\varphi$ is the local longitude of the observer in the body-fixed coordinate system.
  }\label{lcshape}
\end{figure}

The observation geometry at the time of each observation can be easily obtained according
to the orbit of Beagle and WISE. The spin period and spin orientation together with 
shape model of Beagle are also known by light curve inversion method from the DAMIT database. 
The utilized shape model of Beagle is shown in Figure \ref{lcshape}, where $ph$ represents 
rotational phase defined as $ph=1-\varphi/(2\pi)$, in which $\varphi$ is the local longitude of 
the observer in the body-fixed coordinate system.

The scattering weight-factor $w_{\rm f}$ is introduced in the combined Lambert-Lommel-Seeliger law via
\begin{equation}
C_{\rm L}(\psi_i,\psi_{\rm o,i},\alpha,w_{\rm f})=f(\alpha)\left(w_{\rm f}+\frac{1}{\psi_i+\psi_{o,\rm i}}\right),
\label{ScatCoeff}
\end{equation}
where $C_{\rm L}$ represents a correction coefficient to the Lambertian reflection, 
$\psi_i$ and $\psi_{\rm o,i}$ are the cosines of the incident angle and emergence angle
on facet $i$ respectively, $\alpha$ is the solar phase angle; $f(\alpha)$ is the phase correction
function, according to \citep{Kaasalainen2001b},
\[f(\alpha)\sim0.5\exp(-\alpha/0.1)-0.5\alpha+1.\]
Parameter $w_{\rm f}$ represents the weight of Lambertian term in the scattering law, and is 
an artificial factor to interpret sunlight refection. Its physical significance isn't
that clear, thus we only need a scattering weight-factor $w_{\rm f}$ that could achieve 
best-fitting degree to the observations.

The effective diameter $D_{\rm eff}$, defined by the diameter of a sphere with the same area 
to that of the shape model, can be related to the geometric albedo $p_{v}$ and absolute visual 
magnitude $H_{v}$ via:
\begin{equation}
D_{\rm eff}=\frac{1329\times 10^{-H_{v}/5}}{\sqrt{p_{v}}}~(\rm km) ~.
\label{Deff}
\end{equation}
In addition, the geometric albedo $p_{v}$ is related to the effective
Bond albedo $A_{\rm eff,B}$ by
\begin{equation}
A_{\rm eff,B}=p_{v}(0.290+0.684G)~,
\label{aeffpv}
\end{equation}
in which $G$ is the slope parameter in the $H, G$ magnitude system of
\citet{Bowell}. As for Beagle, this work uses $H_{v}=10.8$ and $G=0.15$ (IAU MPC).

So parameters including geometric albedo $p_{\rm v}$, roughness $\theta_{\rm RMS}$, 
and mean grain radius $b$, are the free parameters that would be evaluated from the 
fitting procedures.
\begin{table}[htbp]
 \centering
 \caption{Fixed input parameters of (656) Beagle.}
 \label{inpas}
 \begin{tabular}{@{}ll@{}}
 \hline
 Properties  &  Value \\
 \hline
 Facets of shape model    & 1144 \\
 Nodes of shape model & 574 \\
 Spin ecliptic orientation & ($237^\circ$,$86^\circ$) \\
 Spin period      & 7.0331 hr \\  
 Absolute visual magnitude $H_{v}$  & 10.8 \\
 Slope parameter $G$                & 0.15 \\
 Mean thermal emissivity $\varepsilon$          & 0.9   \\
 Regolith grain density $\rho_{\rm g}$       & $3110~{\rm kgm^{-3}}$ \\
 Regolith porosity $\phi$                       & 0.5   \\
 \hline
\end{tabular}
\end{table}

\subsection{Wavelength-dependent Thermal Emissivity}
\begin{figure}[htbp]
\includegraphics[scale=0.5]{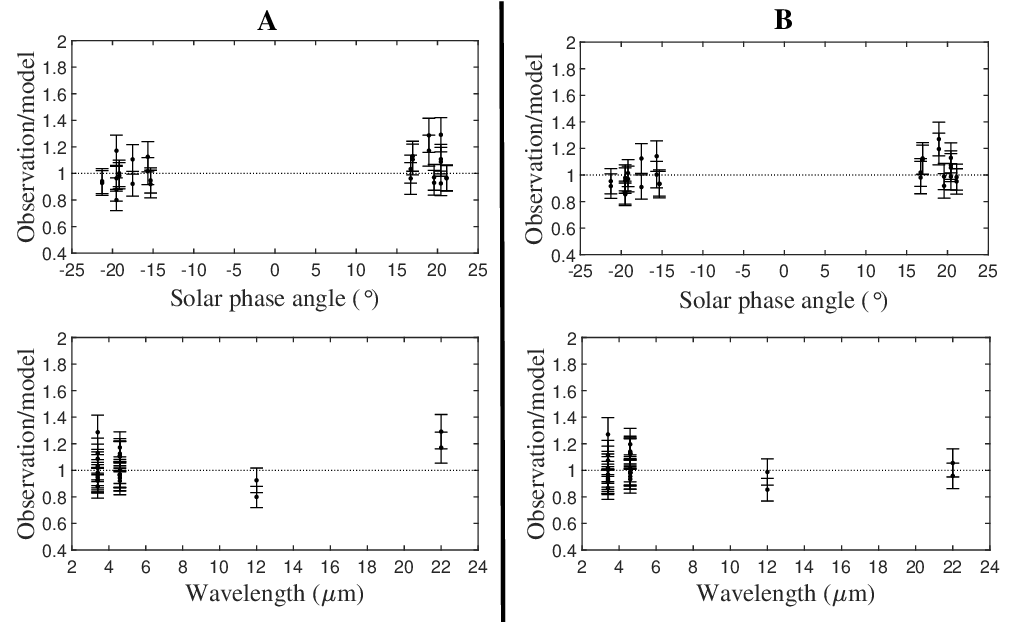}
  \centering
  \caption{The observation/model ratios as a function of wavelength (upper panel) and  solar phase angle (under panel) for the case of best-fit parameters. Panel A (left): best-fit model results are obtained by using constant input thermal emissivity $\epsilon\approx0.9$; Panel B (right): using wavelength-dependent thermal emissivity $\epsilon\approx0.72,0.72,0.72,0.95$ for the W1, W2, W3, W4 band respectively.
  }\label{phwl}
\end{figure}

In general cases, an constant thermal emissivity $\epsilon\approx0.9$ is assumed for each band in the model 
to calculate the thermal emission component. However, in the case of Beagle, we find that, if using 
wavelength-dependent thermal emissivity, the model results match the observations better, as shown in 
Figure \ref{phwl}.

After a process of adjusting the thermal emissivity to fit the observations, we find that, 
the fitting degree can be the best by using $\epsilon\approx0.72,0.72,0.72,0.95$ for the 
W1, W2, W3, W4 band respectively. The thermal emissivity of band W4 is almost close to 1, 
indicating that band W4 is near the Wien peak of the thermal emission spectra of Beagle.

\begin{figure}[htbp]
\includegraphics[scale=0.58]{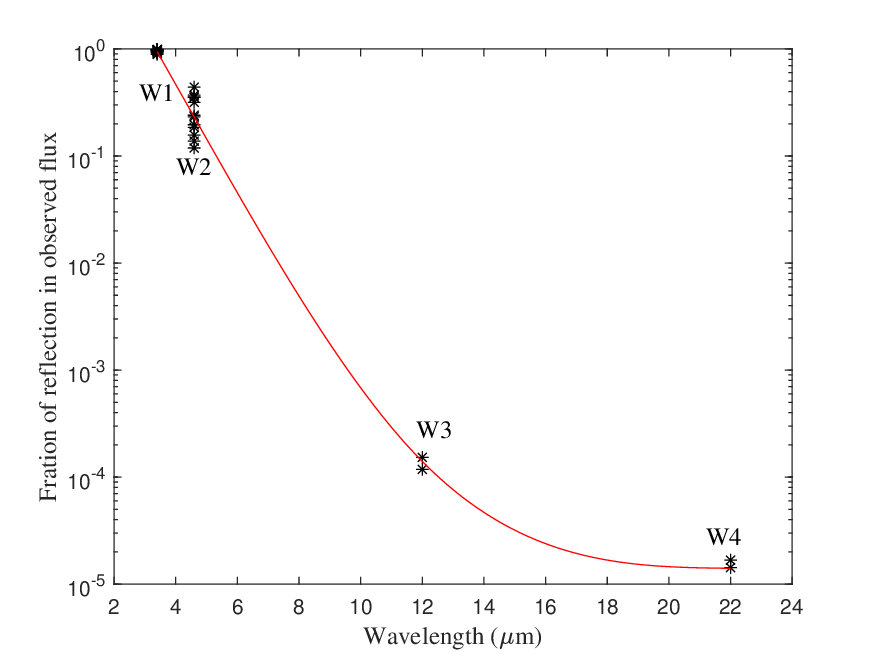}
  \centering
  \caption{The fraction of sunlight-reflection in the observed flux of Beagle for each band of WISE/NEOWISE.
  }\label{rlratio}
\end{figure}

On the other hand, with RSTPM, the fraction of sunlight-reflection in the observed flux of Beagle 
for each band of WISE/NEOWISE can be evaluated, the results are plotted in Figure \ref{rlratio}. 
The W1-band is reflection dominated with reflection ratio $\sim95\%$, the W2-band becomes thermal 
emission dominated but have significant reflection ratio between $10\sim45\%$, finally the W3 and W4 
are almost all thermal emission with reflection ratio $<0.1\%$.

\subsection{Results of Roughness, Grain Size and Albedo}
\begin{figure}[htbp]
\includegraphics[scale=0.58]{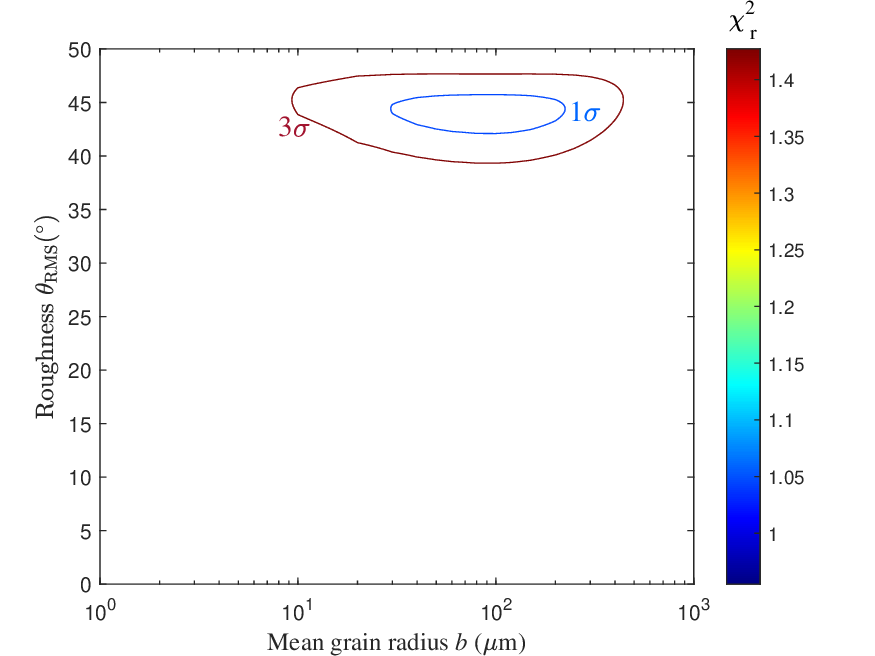}
  \centering
  \caption{The contour of $\chi_{\rm r}^2(\theta_{\rm RMS},b)$, which is obtained by fitting the
observations with two free parameters: the roughness RMS slope $\theta_{\rm RMS}$  and the mean
grain radius $b$.  The blue region stands for the $1\sigma$-level constraint, and the red
region represents the $3\sigma$-level constraint.
  }\label{GaRMS}
\end{figure}

By adopting different thermal emissivity for the four band, we then fit the observations 
by scanning roughness $\theta_{\rm RMS}$ in the range of $0\sim50^\circ$ and mean grain radius 
$b$ in the range of $1\sim1000~\mu$m. With each pair of ($\theta_{\rm RMS}$,$b$), 
a best-fit geometric albedo $p_{\rm v}$ together with effective diameter $D_{\rm eff}$ is found 
to compute the reduced $\chi^2_{\rm r}$. The results are presented in Figure \ref{GaRMS}
as a contour of $\chi^2_{\rm r}$($\theta_{\rm RMS}$,$b$).

According to Figure \ref{GaRMS}, a well constrained $1\sigma$-level limit is derived for 
roughness $\theta_{\rm RMS}$ and mean grain radius $b$, giving $\theta_{\rm RMS}=44\pm2^\circ$, 
$b=100^{+130}_{-70}~\mu$m respectively. $3\sigma$-level constraint for roughness is derived as
$\theta_{\rm RMS}=44\pm4^\circ$, whereas for mean grain radius, a relatively wide 
$3\sigma$-level limit is obtained as $b=100^{+350}_{-90}(10\sim450)~\mu$m.

\begin{figure}[htbp]
\includegraphics[scale=0.58]{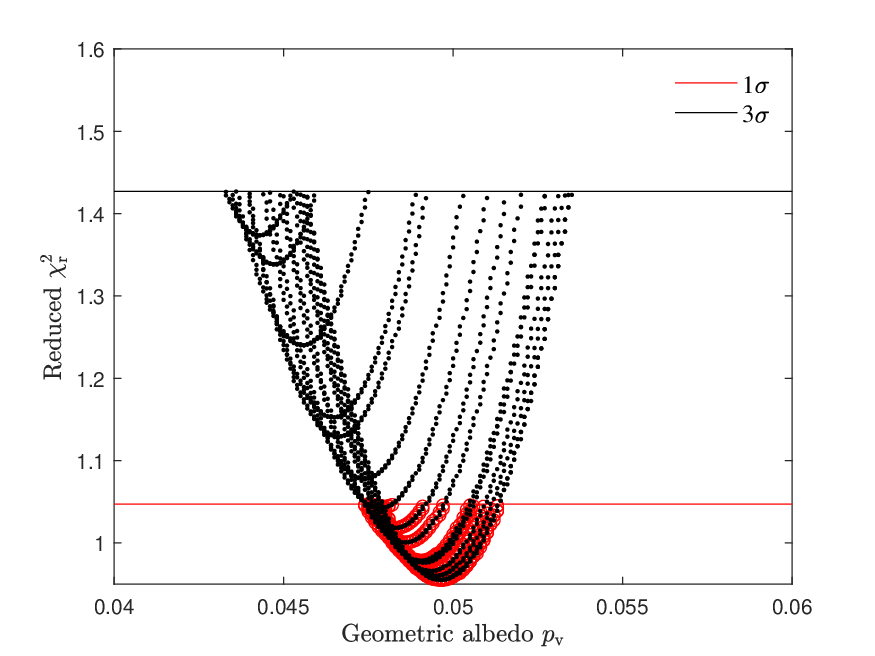}
  \centering
  \caption{$p_{\rm v}\sim\chi^2_{\rm r}$ profiles fit to the observations in consideration of the derived $1\sigma$ and $3\sigma$ ranges of roughness and mean grain radius.
  }\label{pvchi2}
\end{figure}

According to the above derived $1\sigma$ and $3\sigma$ ranges of roughness and mean grain radius, 
the corresponding geometric albedo $p_{\rm v}$ and $\chi^2_{\rm r}$ are picked out, leading to 
the $p_{\rm v}\sim\chi^2_{\rm r}$ relation as shown in Figure \ref{pvchi2}. In this way, we obtain 
the $1\sigma$ and $3\sigma$-level limits of geometric albedo as $p_{\rm v}=0.05^{+0.002}_{-0.003}$ 
and $p_{\rm v}=0.05^{+0.004}_{-0.007}$ respectively, and simultaneously the effective
diameter of Beagle is obtained as $D_{\rm eff}=57.3^{+1.8}_{-1.1}$ km ($1\sigma$) and
$D_{\rm eff}=57.3^{+4.5}_{-2.2}$ km ($3\sigma$).

\subsection{Regolith thermophysical characteristics}
Thermal inertia is a strong function of temperature. Now with the above derived profile of 
mean grain radius, we can evaluate the change of surface thermal inertia of Beagle due to 
the influence of seasonal temperature variation according to the relationships between 
thermal inertia, thermal conductivity, specific heat capacity and temperature 
given in \citet{Yu2021}.

\begin{figure}[htbp]
\includegraphics[scale=0.5]{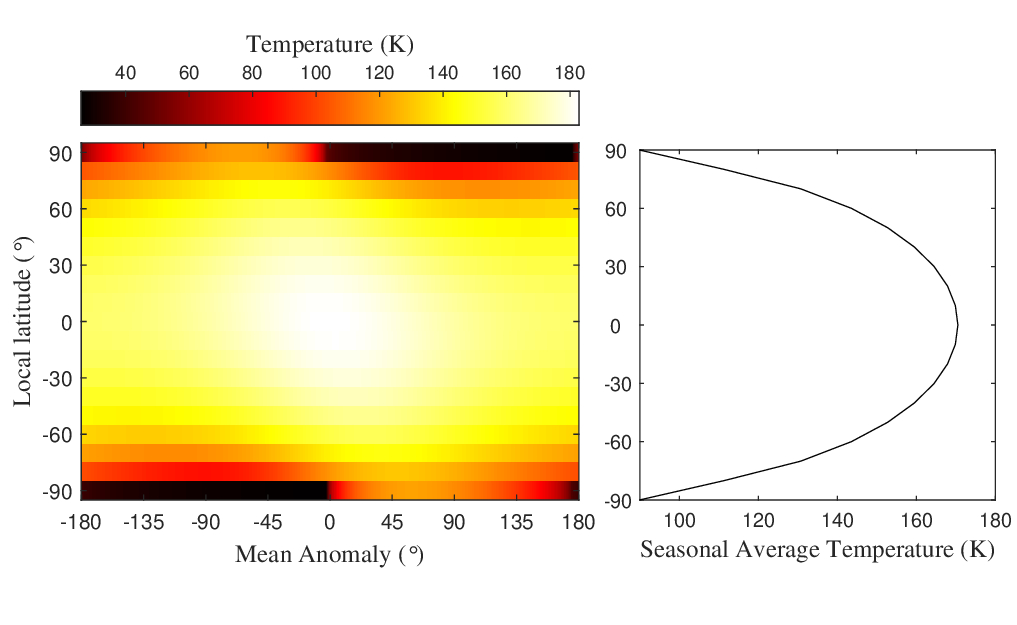}
  \centering
  \caption{Left panel: Seasonal variation of the diurnal-averaged surface temperature as a function of local latitude. Right panel: Seasonal average temperature of each local latitude. The so-called local latitude is defined as the complementary angle of the angle between the local normal vector and the rotation axis.
  }\label{Thsst}
\end{figure}

In the left panel of Figure \ref{Thsst}, a map of surface temperature of Beagle is plotted as 
a function of local latitude and orbital mean anomaly. Each temperature has been averaged over 
one rotational period. We can clearly see that temperature on each local latitude can reach 
maximum (summer) or minimum (winter) at different orbital positions as a result of seasonal 
effect. Temperature on the poles can vary from $\sim$26 K to $\sim$183 K. 
\begin{figure}[htbp]
\includegraphics[scale=0.54]{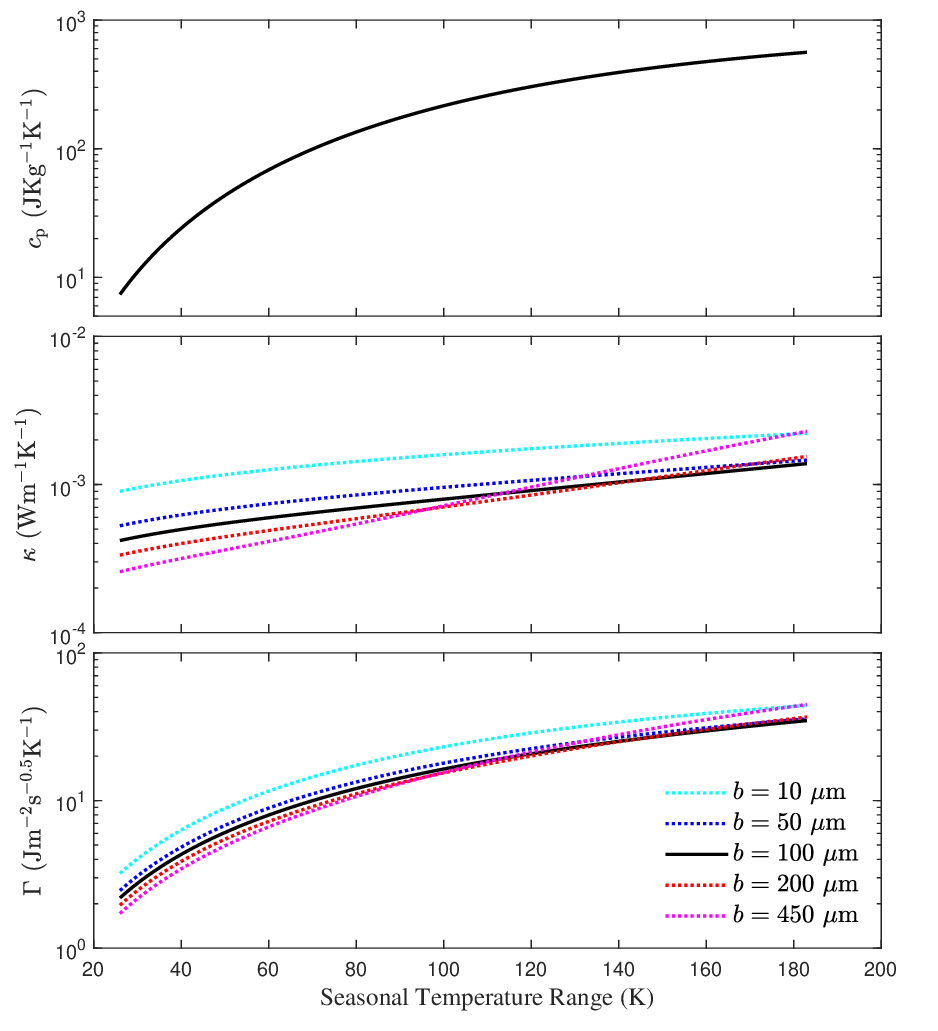}
  \centering
  \caption{The regolith specific heat capacity $c_{\rm p}$, thermal conductivity $\kappa$, and thermal inertia $\Gamma$ of Beagle in consideration of seasonal temperature variation and $3\sigma$-level range of mean grain radius $b$. $\Gamma=\sqrt{(1-\phi)\rho_{\rm g}c_{\rm p}\kappa}$, and porosity $\phi\sim0.5$, regolith grain density $\rho_{\rm g}=3110\rm~kgm^{-3}$.
  }\label{ThTv}
\end{figure}

Figure \ref{ThTv} shows that, if considering seasonal temperature variation and $3\sigma$-level 
range of mean grain radius $b$, the regolith thermal inertia, thermal conductivity, specific 
heat capacity of Beagle can be estimated to vary between $2\sim45\rm~Jm^{-2}s^{-0.5}K^{-1}$, 
$0.3\sim2.3\times10^{-3}\rm~Wm^{-1}K^{-1}$, $7\sim562\rm~JKg^{-1}K^{-1}$.

Despite the fact that thermal parameters is temperature dependent, the majority of relevant 
works ignore such temperature dependence and only estimate the average values. For comparison 
with such existing results, we estimate the seasonal average thermal parameters of Beagle with 
inputs of the derived mean grain radius and seasonal averaged temperature. The seasonal average 
temperature would be a function of local latitude, $\tilde{T}(\theta)$, and can be estimated as
\begin{equation}
(1-A_{\rm eff,B})\tilde{L}_{\rm s}(\theta)=\varepsilon\sigma \tilde{T}(\theta)^4,
\end{equation}
where $A_{\rm eff,B}$ is the bond albedo, $\varepsilon\sim0.9$ is the average thermal
emissivity, $\tilde{L}_{\rm s}(\theta)$ is the annual average incoming solar flux on
each latitude. The results are presented in The right panel of Figure \ref{Thsst}, giving 
the seasonal average temperature of Beagle to be $90\sim170\rm~K$, and accordingly 
the average specific heat capacity 
\[c_{\rm p}=173\sim516\rm~JKg^{-1}K^{-1},\]
the average thermal conductivity 
\[\kappa=0.7\sim1.3\times10^{-3}\rm~Wm^{-1}K^{-1},\]
and the average thermal inertia 
\[\Gamma=\sqrt{(1-\phi)\rho_{\rm g}c_{\rm p}\kappa}=14\sim32\rm~Jm^{-2}s^{-0.5}K^{-1},\]
where the porosity $\phi$ uses 0.5, and the regolith grain density $\rho_{\rm g}$ 
uses $3110\rm~kgm^{-3}$ \citep{Opeil2010}.

\subsection{Infrared light curves}
With the above derived physical and thermophysical parameters of Beagle, now we are able to 
determine the rotation phase of Beagle at each time of WISE/NEOWISE observation by doing 
comparisons between the observational and theoretical light curves. To do so, the 3D shape model 
is used to define the local body-fixed coordinate system, where the z-axis is chosen to be 
the rotation axis. Moreover, if we define the view angle of one observation with respect to 
the body-fixed coordinate system to be $(\varphi,\theta)$, where $\varphi$ stands for local 
longitude, and $\theta$ means local latitude, then the rotational phase $ph$ of this observation 
can be related to the local longitude $\varphi$ via
\begin{equation}
ph=1-\varphi/(2\pi),
\end{equation}
and "zero" rotational phase is chosen to be the "Equatorial view ($0^\circ$)"  
as shown in Figure \ref{lcshape}.

If selecting a reference epoch, and assuming the rotational phase at this epoch to be
$zph$, then all the rotational phases of other data could be derived in consideration
of the observation time and geometry. Furthermore, for some particular epoch, light curves 
can be derived for each band by correcting the observed flux at various epochs into one 
rotation period at this epoch, where the correction is implemented via
\begin{equation}
F_{i,\rm corr}=F_i\left(\frac{r_{i,\rm helio}}{r_{0,\rm helio}}\right)^2
\left(\frac{\Delta_{i,\rm obs}}{\Delta_{0,\rm obs}}\right)^2,
\label{fluxcorr}
\end{equation}
in which $F_{i,\rm corr}$ is the flux after correction, $F_i$ is the original observed flux,
$r_{i,\rm helio}$ and $r_{0,\rm helio}$ are the heliocentric distance of epoch $i$ and the
reference epoch, while $\Delta_{i,\rm obs}$ and $\Delta_{0,\rm obs}$ are the observation distance. 

\begin{figure*}[htbp]
\includegraphics[scale=0.48]{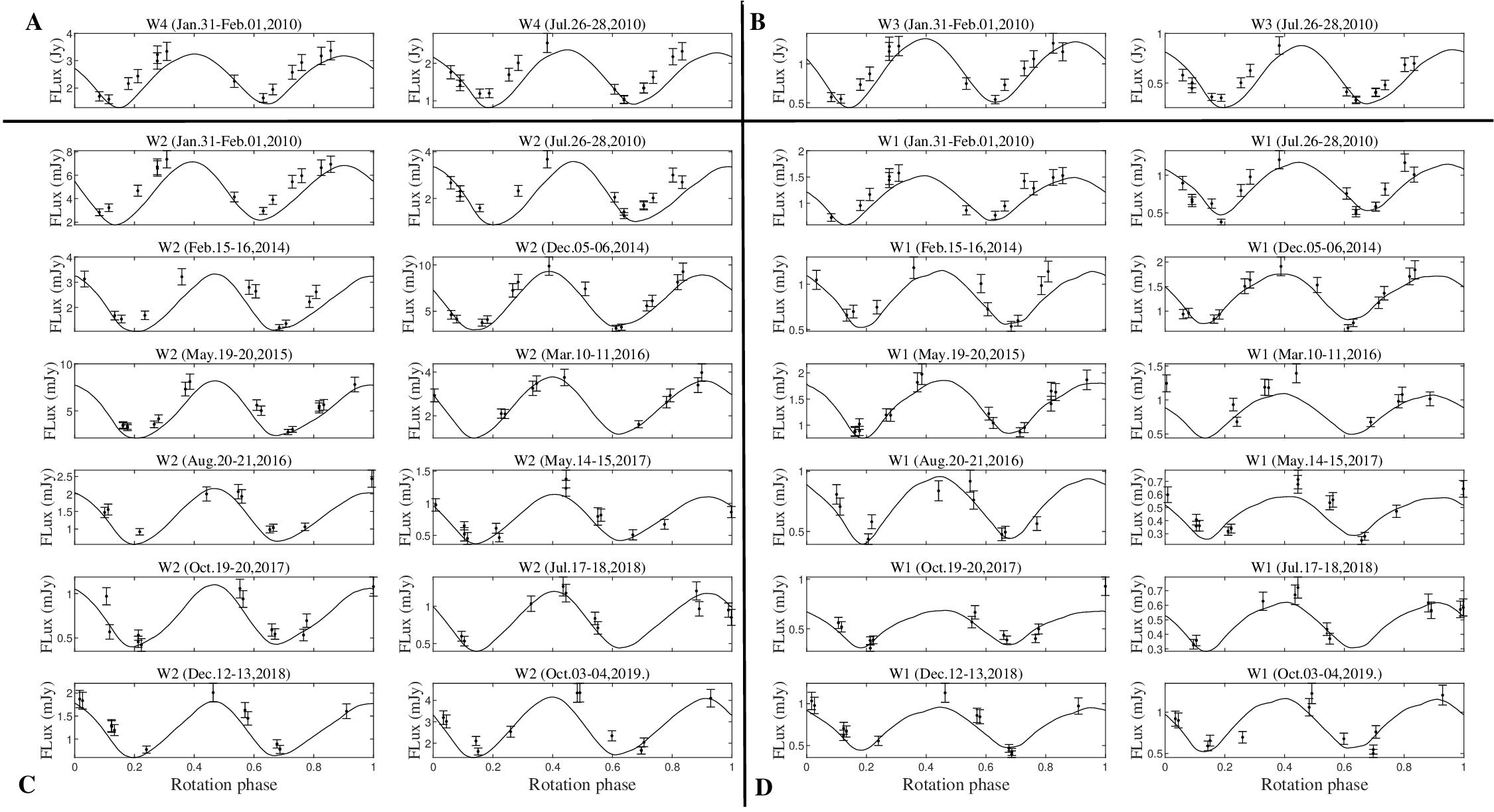}
  \centering
  \caption{Best-fit results to the infrared light curves of WISE/NEOWISE for each observation epoch; Panel A,B,C: thermal light curves at band W4, W3, W2 respectively; Panel D: reflection light curves at band W1.
  }\label{lcW1234}
\end{figure*}

Following the above method, firstly we select '2010-01-31 12:06' as the reference epoch for
deriving the reference rotational phase $zph$. But for correction of flux, in order to reduce
flux errors caused by correction Equation (\ref{fluxcorr}), we select 28 separate reference epochs,
so as to use data close to each reference epoch (data within three days) to generate infrared 
light curves. By this way, we obtain 28 infrared light curves. Then for each of the reference epoch, 
theoretical infrared light curves are simulated by RSTPM to fit the above generated observational 
light curves. The best-fit results are plotted in Figure \ref{lcW1234}. Roughly speaking, our modeled 
infrared light curves match well with all the four-band observational light curves of WISE/NEOWISE 
in consideration of the observation errors, indicating that our model results are reliable.
\begin{figure*}[htbp]
\includegraphics[scale=0.78]{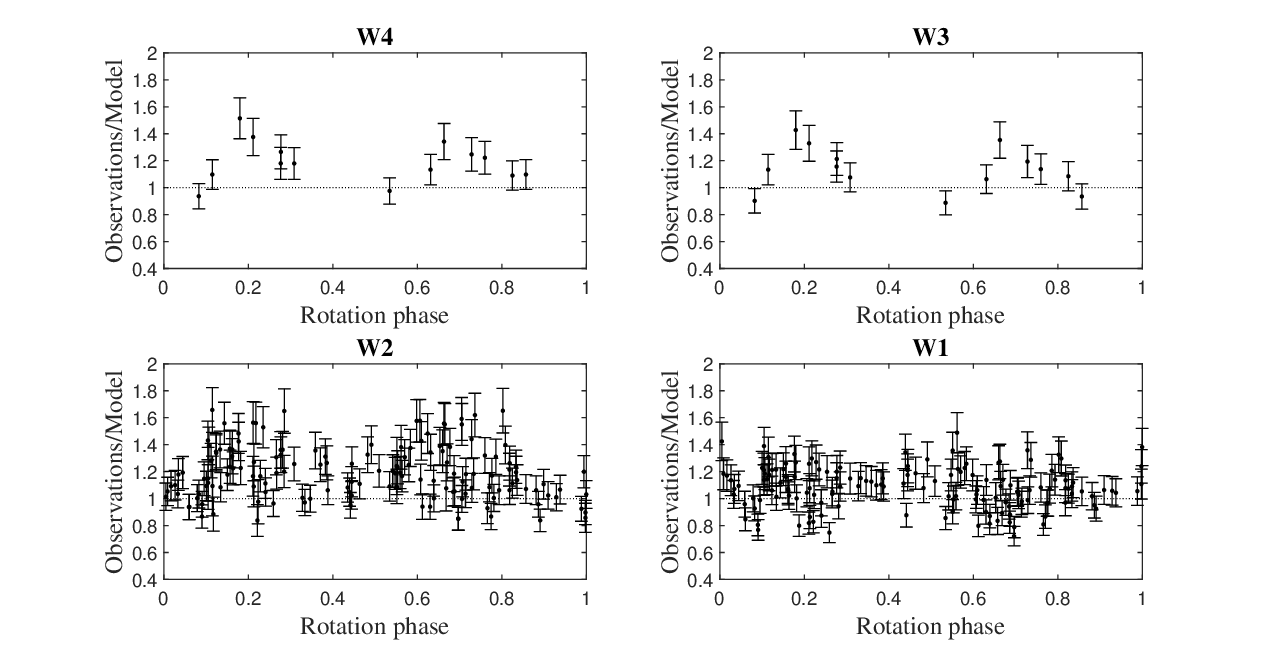}
  \centering
  \caption{Ratios of observation/model as a function of rotation phase for each band respectively.
  }\label{ntlc}
\end{figure*}

To investigate whether Beagle has surface heterogeneity, the ratios of observation/model are plotted 
as a function of rotation phase for each band respectively in Figure \ref{ntlc}, where a clear rotation 
phase dependent feature is observed in the data of band-W4, W3, and W2, showing that the observation/model 
ratios appear to be significantly higher around $ph=0.2$ and 0.7, whereas the band-W1 data don't show 
such feature. Since both irregular shape and surface heterogeneity can contribute to the rotational 
variation of light-curves, and bands of W4, W3, and W2 are thermal emission dominated, so the 
heterogeneous features of the thermal light curves of WISE/NEOWISE may imply that: 1, the light-curve 
inversion shape model is not perfect for modeling thermal emission; or 2,  the regolith of Beagle may 
have heterogeneous thermophysical characteristics along longitude.

\section{Discussion}
The multi-epoch infrared data of WISE/NEOWISE make it possible to probe the regolith 
characteristics of asteroids in detail. As an example, this work uses 9 years of infrared data from 
WISE/NEOWISE to study the regolith of Main-Belt Object (656) Beagle by the well-tested thermophysical 
model --- RSTPM \citep{Yu2021}. Details about the data processing, fitting procedure, modelling and 
analyzing of infrared light curves are presented. The results show that Main-Belt Object Beagle 
has a low mean thermal inertia of $\Gamma=14\sim32\rm~Jm^{-2}s^{-0.5}K^{-1}$, being consistent with 
that of most large main belt objects \citep{MacLennan2021,MacLennan2022,Huang2022}; and Beagle 
has a geometric visual albedo as low as $0.043\sim0.054$, being far lower than the geometric 
albedos of its neighbouring asteroids (mean $p_{\rm v}=0.0941\pm0.0055$) \citep{Fornasier2016}, even 
lower than the geometric albedo of (24) Themis ($p_{\rm v}=0.064^{+0.008}_{-0.011}$) \citep{Yu2021}. 
Although the low albedo $\sim0.05$ of Beagle still lies on the edge of $1\sigma$ range of the 
Beagle family and the Themis family, it is a little anomalous that the albedos of Beagle's neighbouring 
asteroids are more close to Themis, rather than Beagle itself.

In addition, Figure \ref{ntlc} shows that Beagle exhibits significant rotation-phase dependent features at 
band-W4, W3 and W2, but not at W1-band. According to Figure \ref{rlratio}, band-W4, W3 and W2 are thermal 
emission dominated, whereas W1-band is dominated by NIR reflection. So the result implies that the surface 
of Beagle doesn't have significant heterogeneous NIR reflectivity. As mentioned before, \citet{Fornasier2016} 
shows that the neighbouring asteroids of Beagle have diverse NIR spectral types, and Beagle is expected 
to show heterogeneous NIR features across its surface if Beagle is the parent of those neighbouring asteroids. 
However, here our result shows that Beagle has no significant heterogeneous NIR reflectivity, indicating that 
Beagle may be not the parent of its neighbouring asteroids, or surface heterogeneity of Beagle at NIR are 
eliminated for some reasons, for example, the utilization of imperfect light-curve inversion shape model, 
which doesn't remove the degeneracy between regional NIR reflectivity and slope.

On the other hand, for band-W4, W3 and W2, if doing comparisons between Figure \ref{ntlc} and \ref{lcshape}, 
we can find that the higher observation/model ratios appear around $ph=0.2$ and 0.7, the phases of which 
Beagle happens to have the minimum cross-section area, implying that the rotation-phase dependent feature 
of band-W4, W3 and W2 has a strong relationship with the utilized shape model. Also we have done several 
tests by adjusting the rotation period within $7.033\pm0.001$ to do the fitting procedure, but the above 
rotation-phase dependent feature remains unchange, indicating that the utilized shape rather than rotation 
period is more likely to be cause of the rotation-phase dependent feature. Hence the light-curve inversion 
shape model from optical light curves is not perfect for infrared light curves, similar phenomenons have 
been found by \citet{Hanus2015} and \citet{Durech2017}. Nevertheless, the possibility of heterogeneous 
regolith thermophysical characteristics across the surface of Beagle remains positive, because it would 
be reasonable to expect different physical characteristics (e.g. roughness) between the body part and the 
smaller head part, as such phenomenons have been observed on other asteroids by in-situ space missions, 
for example (25143) Itokawa \citep{Abe2006}.

\section{Conclusion}
By analyzing 9-years infrared light curves of Beagle from WISE/NEOWISE, we get the following results: 

1). Beagle has an effective diameter $D_{\rm eff}=57.3^{+4.5}_{-2.2}$ km, geometric albedo 
$p_{\rm v}=0.05^{+0.004}_{-0.007}$, mean roughness $\theta_{\rm RMS}=44\pm4^\circ$, mean grain 
size $b=100^{+350}_{-90}(10\sim450)~\mu$m, mean specific heat capacity $c_{\rm p}=173\sim516\rm~JKg^{-1}K^{-1}$, 
thermal conductivity $\kappa=0.7\sim1.3\times10^{-3}\rm~Wm^{-1}K^{-1}$ and thermal inertia 
$\Gamma=14\sim32\rm~Jm^{-2}s^{-0.5}K^{-1}$. 

2). We confirm that the Beagle has an anomalous low albedo that the albedos of Beagle's neighbouring 
asteroids are more close to Themis, rather than Beagle itself. In addition, the W1-band NIR light 
curves of Beagle don't reveal significant heterogeneous NIR reflectivity across the surface of Beagle. 
This result does not support Beagle to be the parent of its neighbouring asteroids which have diverse 
NIR spectral types \citep{Fornasier2016}. A possible explanation to these results is that Beagle 
may be an interloper or a sister, rather than the parent of its neighbouring asteroids including 
the first MBC 133P. Considering Beagle's albedo anomaly, we prefer to surmise Beagle to be 
an interloper to the Themis family, although it still remains possible for Beagle being an anomalous 
member of the Themis family. So our results may add new clues of Beagle probably having no genetic 
connection with its neighbouring asteroids and even (24) Themis, and may lead to new scenarios about 
the origin of the famous MBC 133P.

3). Asteroidal shape models from inversion of optical light curves could have obvious imperfections.  
As a result, the heterogeneity of surface reflectivity at near infrared is difficult to be discovered 
if using such shapes to model the theoretical infrared light curves. Besides, the imperfect shapes 
can also lead to a rotation dependent features in thermal light curves (e.g. band W2, W3, W4 of WISE 
and NEOWISE), which would mislead the evaluation of the heterogeneity of regolith thermophysical 
characteristics. Therefore, to obtain truthful information about the surface heterogeneity of asteroids, 
it is necessary to input shapes that are more close to reality.

\section*{Acknowledgments}
We would like to thank the WISE teams for providing public data. This work was supported by the grants from The Science and Technology Development Fund, Macau SAR (File No. 0051/2021/A1). and Faculty Research Grants of The Macau University of Science and Technology (File).

\bibliographystyle{named}

\end{document}